\documentclass[12pt,preprint]{aastex}
\usepackage{natbib}
\usepackage{booktabs}
\usepackage{graphicx}
\usepackage{amsmath,amssymb}

\shortauthors{T. J. Veach}
\shorttitle{Vibrationally Excited HCN around AFGL 2591}

\begin{document}

\title{Vibrationally Excited HCN around AFGL 2591: A Probe of Protostellar Structure }

\author{Todd J. Veach}
\email{tveach@asu.edu}
\affil{Department of Physics, Arizona State University, Tempe, AZ 85287}

\author{Christopher E. Groppi}
\affil{School for Earth and Space Exploration, Arizona State University, Tempe, AZ 85287}

\author{Abigail Hedden}
\affil{ Harvard-Smithsonian  Center for Astrophysics, 60 Garden Street, MS-78, Cambridge, MA 02138}

\begin{abstract}
Vibrationally excited molecules with sub-mm rotational transitions are potentially excellent probes of physical conditions near protostars. This study uses observations of the $v=1$ and $v=2$ ro-vibrational modes of HCN (4-3) to probe this environment. The presence or absence and relative strengths of these ro-vibrational lines probe the gas excitation mechanism and physical conditions in warm, dense material associated with protostellar disks. We present pilot observations from the Heinrich Hertz Submillimeter Telescope (HHSMT) and followup observations from the Submillimeter Array (SMA). All vibrationally excited HCN (4-3) $v=0$, $v=1$, and $v=2$ lines were observed. The existence of the three $v=2$ lines at approximately equal intensity imply collisional excitation with a density of greater than ($ 10^{10}$ cm$^{-3}$) and a temperature of  $>1000$K for the emitting gas. This warm, high density material should directly trace structures formed in the protostellar envelope and disk environment. Further, the line shapes of the $v=2$ emission may suggest a Keplerian disk. This letter demonstrates the utility of this technique which is of particular interest due to the recent inauguration of the Atacama Large Millimeter Array (ALMA). 

\end{abstract}

\keywords{ISM: molecules ---  stars: formation --- stars: individual (AFGL 2591) --- stars: kinematics and dynamics --- stars: protostars --- submillimeter: stars}

\section{Introduction} \label {intro}

The vast majority of the molecular lines observed at millimeter wavelengths are associated with the lowest vibrational state ($v = 0$). However, several ro-vibrational molecular transitions have been detected in the ISM and towards hot molecular cloud cores (CS, H$_{2}$, SiO, HC$_{3}$N, CH$_{3}$CN, and HCN \citep{1986ApJ...300L..19Z,1987A&A...182L..15T,2011A&A...529A..76R,2011A&A...536A..33R}). For vibrational excitation to occur these molecules must be in regions of hot, dense gas (perhaps associated with a shock) and/or pumped by infrared radiation. Since these excitation requirements are expected to be fulfilled only nearby to an embedded source, they make vibrationally excited molecules with sub-mm/mm wave rotational transitions potentially excellent probes of the physical conditions very near embedded infrared sources. This is especially true for those sources which suffer so much extinction that they are unobservable in the infrared \citep{1994ApJ...437L.127W}.

Initially, vibrationally excited HCN was used as a probe of physical conditions for astrochemistry studies \citep{1971ApJ...170L.109M, 1973ApJ...183..871W, 1975ApJ...197..603M}. In \citeyear{1994ApJ...437L.127W}, \citeauthor{1994ApJ...437L.127W} used the fact that the critical density and energy above ground for the $v=1$ vibrationally excited state of CS ($7-6$) and ($10-9$) was high (n$\sim10^8$ cm$^{-3}$ and 2000 K respectively) in order to probe protostellar disk structure toward IRAS 16293, pioneering this technique. An excitation analysis of the vibrational lines suggested emission arising from shocks within a hot, dense circumstellar disk perhaps due to self-gravitating disk instabilities, e.g., \citet{1989ApJ...347..959A}. Since this hot, dense gas could be potentially associated with shocks and/or spiral density waves in a protostellar accretion disk, vibrational molecular transitions could be used as a probe of disk structure and planet formation. 

We have chosen to observe the $v=0$, $v=1$ and $v=2$ vibrational modes of HCN (4-3) rather than CS. The critical densities of HCN are even higher than CS ($> 10^{10} $\ cm$^{-3}$), but the excitation temperature of the $v=1$ ro-vibrational transitions is $\sim 1050$ K, half that of CS. This could make HCN an even better probe of dense material in protostellar accretion disks, probing material farther from the protostellar object. For gas at the temperatures necessary to collisionally excite these ro-vibrational states, we expect higher J-levels are preferentially populated \citep{1994ApJ...437L.127W}. Thus, the compromise between atmospheric transparency and line brightness makes HCN (4-3) the best choice for use as a probe of protostellar structure rather than lower lying rotational transitions. 

The selection rules governing the $v=1$ and $v=2$ transitions help with observing efficiency and interpretation. Emission from the $(0, 2^{2c}, 0)$ and $(0, 2^{2d}, 0)$ lines is expected to be absent due to $\Delta l = \pm1$ selection rules if purely radiative excitation dominates \citep{1986ApJ...300L..19Z}. If collisional excitation of the $v=1$ and $v=2$ lines is the dominant emission mechanism, we would expect to observe both $v=1$ lines at approximately equal intensity, and all three $v=2$ lines also at approximately equal intensity. Furthermore, the extremely high critical density and temperature of these lines allow observation of emission that would ordinarily be contaminated by foreground and background emission within the beam \citep{1994ApJ...437L.127W}. With contamination from the envelope removed, interpretation of the morphology and kinematics of protostellar sources should be simplified.

AFGL 2591 is a testbed of high-mass star formation that has unusually high HCN abundance ($\sim10^{-6}$, \cite{2001ApJ...553L..63B}). According to \cite{2007A&A...475..549B} and references therein, it can be also be classified as a high mass protostellar object (HMPO) or early ``Hot Core." HMPOs are distinctive stages of star formation in which the sources are bright in the mid-infrared (10-12$\mu$m) and show a central density concentration.  In this phase, the protostellar object is still deeply embedded in a collapsing dusty envelope. These high-mass objects can effectively heat their envelope by FUV radiation. High energy radiation from UV photons or X-rays are unobservable directly, but the effects on the molecules within the surrounding environment can be observed in the submillimeter. \cite{2006A&A...447.1011V} provide interferometric observations of the H$^{18}_{2}$0 line at 203 GHz showing the existence of a rotating flattened structure, which could be evidence for an accretion disk \citep{2010A&A...521L..22C}. The central mass of AFGL 2591 is assumed to between $\sim10 - 16$ M$_{\odot}$ \citep{2000ApJ...541L..63M},\citep{2006A&A...447.1011V}, and the distance to AFGL 2591 is $1\pm1.0/0.5$ kpc \citep{1999ApJ...522..991V}. We adopt a distance of 1~kpc in this Letter for all analyses.

This study makes use of submillimeter interferometric observations from the SMA to reduce the effects of beam dilution and increase sensitivity over single dish observations. The first goal of this study is to determine the primary excitation mechanism for the vibrational emission: collisional excitation or IR pumping. If the emission arises from collisional excitation, it should trace extremely dense, warm gas of the type expected in accretion disk structure. Furthermore, we would expect the emission region to be compact, even at interferometric resolution if produced by dense gas tracing protostellar disk and/or envelope structure. The vastly expanded capabilities of the Atacama Large Millimeter Array (ALMA) will further increase spatial resolution and sensitivity to permit \textit{imaging} of a wide variety of sources in these ro-vibrational transitions with greatly reduced integration time when compared to other observatories. The observations of ALGL2591 presented here demonstrate the utility of this technique, which will be even more effective and efficient when applied with ALMA. 

\section{Observations}\label{ob_red}

\subsection{HHSMT HCN J=4$\rightarrow$3 Observations \& Data Analysis}
Observations of the $v=0$ and $v=(0,1^{1c},0)$  lines for AFGL 2591 ($\alpha=20^{^{h}} 29^{^{m}} 24.90^{^{s}},\delta=40^{^{\circ}} 11^{^{\prime}} 21.0^{^{\prime \prime}} $) were made at the HHSMT in February 2000 during priority observing time. The receiver was tuned to 354.46033~GHz upper sideband (USB). The average T$_{\text{sys}}$ was 528~K. The on-source integration time was 48 minutes. A second tuning covering the $v=(0,1^{1d},0)$ and $v=2$ lines was made in December 2002. The receiver was tuned to 356.25571~GHz USB. The average T$_{\text{sys}}$ for this tuning was 1098~K. The on-source integration time was 93 minutes. The beam size was $\sim$22\arcsec\ for both tunings. Pointing accuracy was $2^{^{\prime \prime}}$ and was checked using planets and strong continuum sources several times per observing session. Since the observations were taken during priority observing time, $\tau_{225GHz} \le 0.08$ for all observations. For both tunings, we used the MPIfR 345 GHz dual polarization SIS receiver, with 1~GHz bandwidth Acousto-Optical Spectrometers providing 1.0~MHz (0.87~$km\ s^{-1}$) spectral resolution. Main beam efficiency of this receiver at 350~GHz is $\eta_{MB} \sim 0.45$ for unresolved sources, computed via total power measurements of Mars following \cite{1993PASP...105..683}. Calibration was performed using the Hot-Sky-Cold method. The wobble switching observing mode was used with a 120\arcsec\ chopper throw, and a 50\% duty cycle. We subtracted linear fits to the baselines for each spectrum, then co-added the spectra, weighted by RMS noise. The final, co-added noise levels in the 354.46033~GHz and 356.25571~GHz tunings were 0.010~K and 0.014~K, respectively. Reductions were done using the GILDAS\footnote{http://www.iram.fr/IRAMFR/GILDAS} software package. 

\subsection{SMA Observations \& Data Analysis}
Observations of the vibrationally excited spectral lines of AFGL 2591 ($\alpha = 20^{^{h}} 29^{^{m}} 24.90^{^{s}},\newline \delta = 40^{^{\circ}} 11^{^{\prime}} 21.0^{^{\prime \prime}} $) $v=0$, $v=(0,1^{1c},0)$, (0, $1^{1d}$, 0), and $v=$(0, $2^0$, 0), (0, $2^{2c}$, 0) and (0, $2^{2d}$, 0) were observed in a single tuning (354.505476 GHz USB) of the SMA in the \textit{subcompact} configuration on September 8, 2008. Our tuning was specifically chosen to place all six of the vibrationally excited HCN lines in the USB. The lower side band (LSB) also included CO(3$\rightarrow$2) and H$^{13}$CN, but these lines were not the focus of this study. The brightest calibrators (mwc349a and 2015+371) closest to the source were used as phase calibrators. Uranus served as flux calibrator, and the quasar 3c454.3 was used for passband calibration.  The precipitable water vapor (PWV) provided in the SMA data headers during observations was less than 2.5 mm over the 7.2 hours of observations. The field of view of the SMA primary beam is 36\arcsec\ at this frequency. Main beam efficiency of the SMA at 354 GHz is $\sim$0.74 for unresolved sources\footnote{sma1.sma.hawaii.edu/beamcalc.html}.

\begin{figure}[!ht] 
   \includegraphics[width=\textwidth]{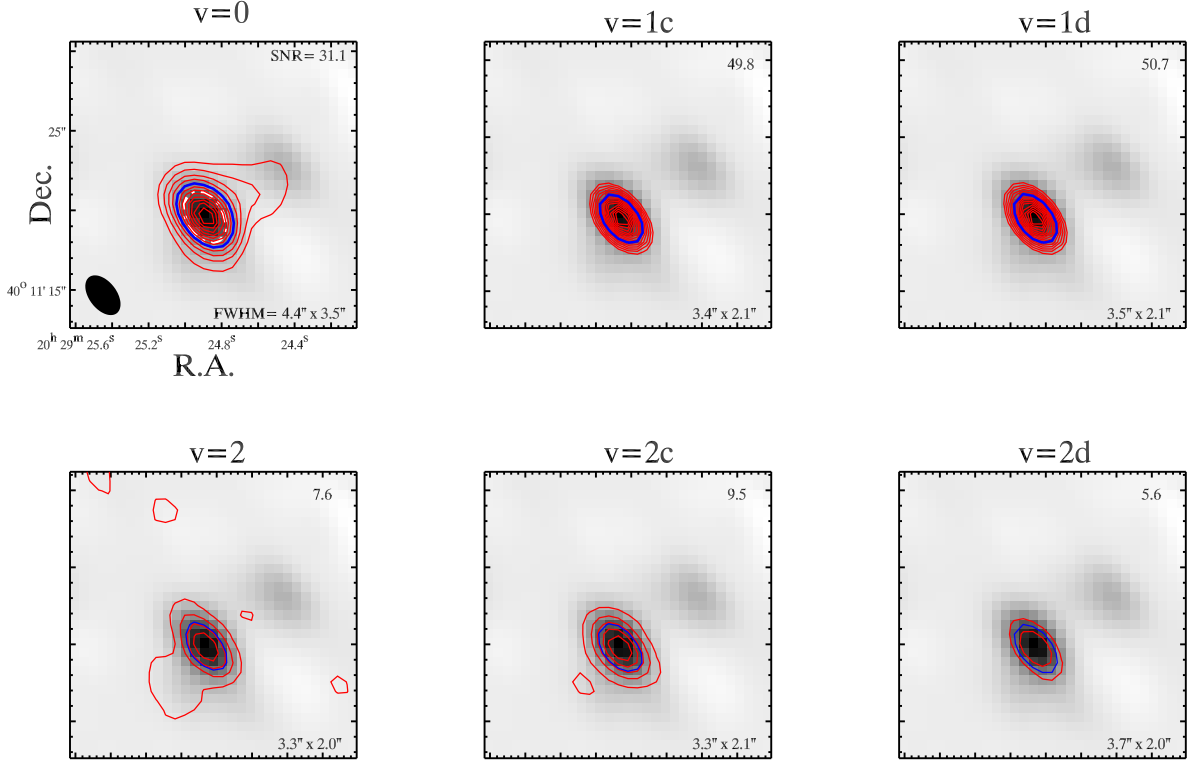} 
   \caption{Integrated intensity contours of HCN emission overlaid on the continuum image in greyscale. The peak flux in the continuum is 1.42~Jy beam$^{-1}$ with an rms noise of 3.4~mJy beam$^{-1}$. The spectra were extracted from the \textit{white} dashed ellipse shown in the upper left panel. The rms map noise is 5.44, 0.92, 0.89, 0.81, 0.82, and 0.75 $K\ km\ s^{-1}$ from top left to bottom right. The top right of each panel is the SNR of the integrated intensity maps, defined as (half the maximum value from a gaussian fit to the integrated intensity (HM))/(the background noise in the integrated intensity image).  The 3.4\arcsec $ \times$ 2.0\arcsec\ synthesized beam is the solid ellipse. The \textit{red} contour levels of the top three panels begin and are spaced at $5\sigma$ intervals of the rms map noise. The \textit{red} contour levels of the bottom three panels begin and are spaced at $3\sigma$ intervals of the rms map noise. The lower right hand corner of each panel shows the size of the FWHM fit to the integrated intensity, also shown as a \textit{blue} contour.  }
   \label{fig:ff1}
\end{figure}

Image reconstruction was done using the Multichannel Image Reconstruction, Image Analysis, and Display (MIRIAD) \citep{1995ASPC...77..433S} software package with extensions for the SMA using the procedures outlined in \citet{Zhao:2009sma}.  The continuum image was constructed using molecular line-free ``chunks'' of the correlator. After molecular line subtraction, the equivalent bandwidth for the continuum was 1.3 GHz. In order to generate a continuum image, we used the MIRIAD INVERT task with uniform weighting, we generated a ``dirty image'' and performed a hybrid Hogbom/Clark CLEAN using a gain of 0.08 with 35000 iterations to produce a cleaned image. The number of iterations used for cleaning was determined systematically by maximizing the map SNR. The ``clean'' maps were then RESTORed by convolution with a gaussian beam, and were then exported to IDL\footnote{IDL, ENVI, and Image Access Solutions (IAS) are trademarks of Exelis Corporation} for further analysis. The FWHM of the synthesized beam is $3.4\arcsec \times 2.0\arcsec $ ($3400 \times 2000$ AU) with a position angle of 37.4$^\circ$. 

CLEANed three-dimensional image cubes, generated similarly to the process described above, were exported to IDL for spectral line extraction. Continuum emission was subtracted in the UV data before the clean and invert steps. The spectral line extraction region is defined by a two-dimensional gaussian fit to the CLEANed continuum image. The FWHM of the extraction region is 4.0\arcsec$\times$2.6\arcsec\ with a 37.4\arcdeg\ rotation angle east of north. No smoothing was applied to the v=0, $v=(0,1^{1c},0)$, and $v=(0,1^{1d},0)$ spectral lines, with a spectral resolution of 0.273 $km\ s^{-1}$. The $v=(0,2^{0},0)$, $v=(0,2^{2c},0)$,$v=(0,2^{2d},0)$ spectra were boxcar smoothed by 3. The resulting continuum and HCN (4-3) integrated intensity maps and associated RMS noise levels are shown in Figure \ref{fig:ff1}.

\section{Results}\label{results}

\begin{figure}[!ht]
\centering
\includegraphics[width=\textwidth]{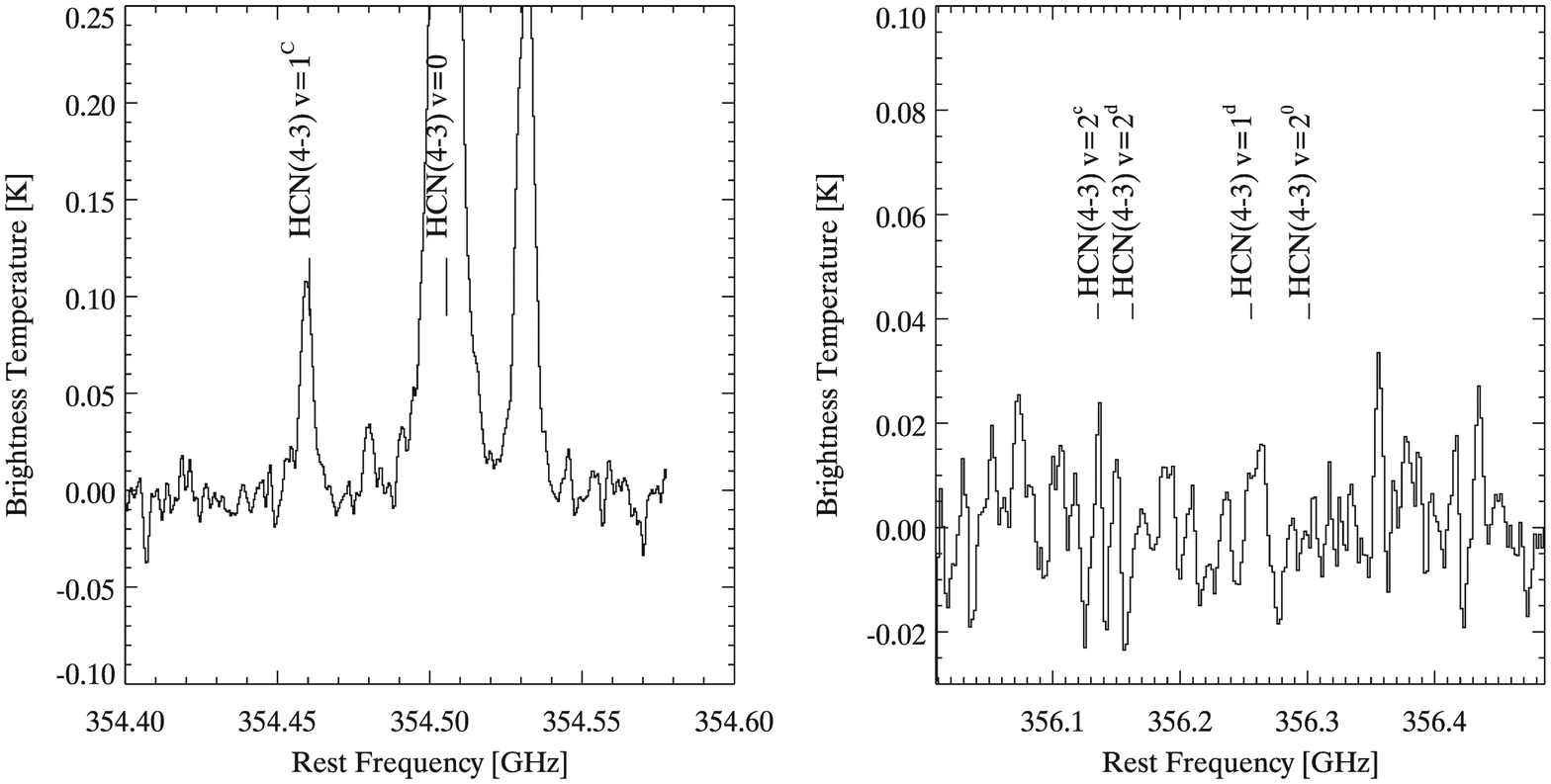}
\caption{Spectra of AFGL 2591 observed with the HHSMT.}
\label{fig:hht1}
\end{figure} 

\subsection{HHSMT Spectra}\label{hht_results}
Spectra of AFGL 2591 observed with the HHSMT are shown in Figure \ref{fig:hht1}. We find bright $v=0$ HCN (4-3) emission and prominent HCN (4-3) $v=(0,1^{1c},0)$ ($\sim110$ mK) emission. However, we find a tentative detection of $v=1d$ line (integrated intensity SNR of 7.6, $1.5\sigma$) to be $5\times$ weaker than expected (i.e. equal in intensity to the $v=1c$ line). We find no $v=2$ lines in the spectrum. Line parameters and upper limits are given in Table 1.

\subsection{SMA Data}\label{sma_results}

The $v=0$ integrated intensity map in panel 1 of Figure \ref{fig:ff1} shows evidence for an extended source towards the northwest of the image, consistent with results from \cite{2007A&A...475..549B}. The $v=(0,1^{1c},0)$ and $v=(0,1^{1d},0)$ maps show a marginally resolved source.  The $v = (0, 2^0, 0)$, $(0, 2^{2c}, 0)$ and $(0, 2^{2d}, 0)$ maps show mostly unresolved  sources, with some hint of extension in the first $3\sigma$ contour level to the southeast in the $v = (0, 2^0, 0)$ line. 

\begin{figure}[!ht]
\centering
{\includegraphics[width=\textwidth]{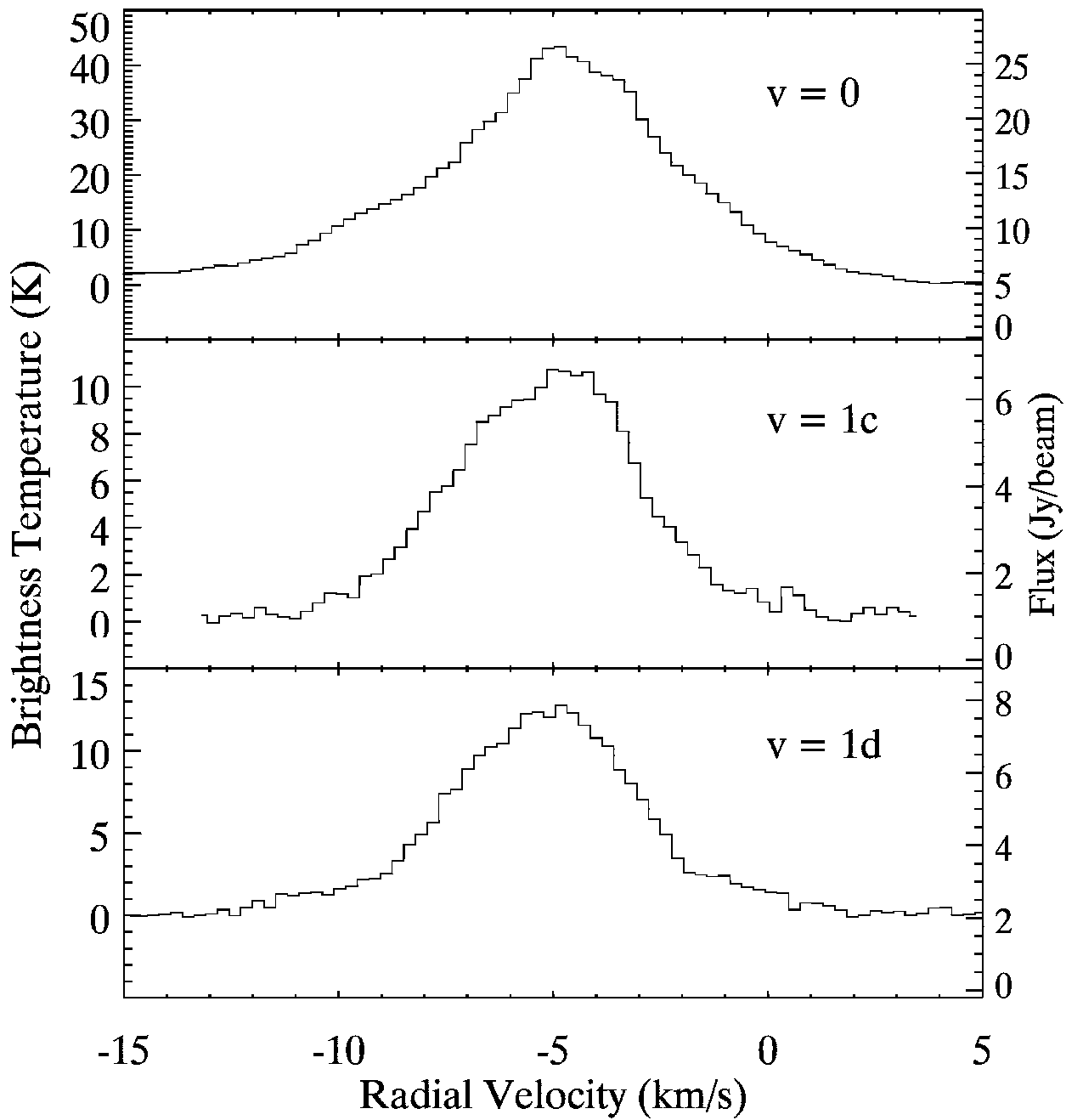}}
\caption{HCN v=0, v=(0,$1^{1c}$,0), and v=(0,$1^{1d}$,0) lines. Velocity resolution is 0.273 $km\ s^{-1}$. }
\label{fig:sma1}
\end{figure}

\begin{figure}[!ht]
\centering
{\includegraphics[width=\textwidth]{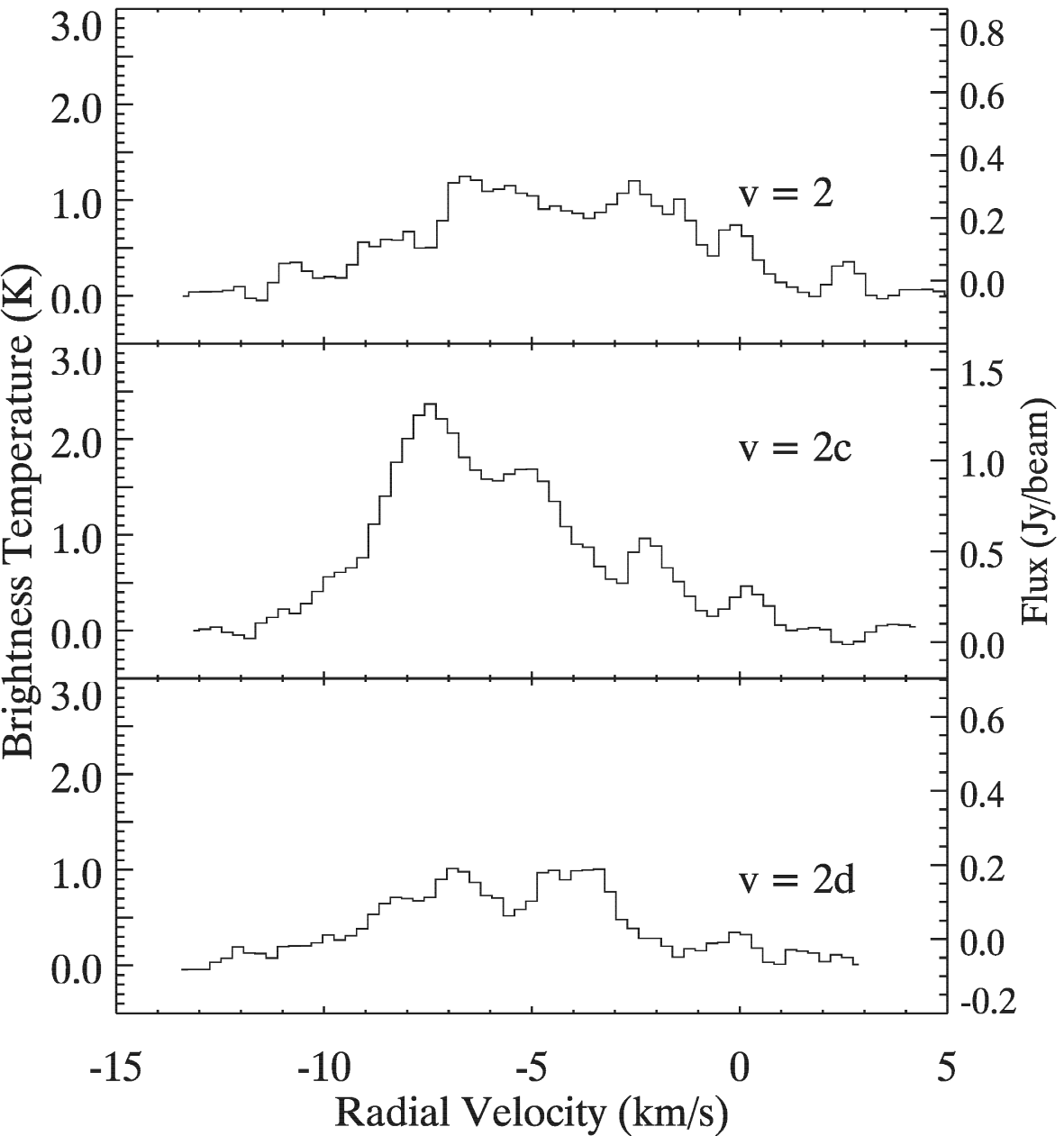}}
\caption{HCN $v=(0,2^{0},0)$, $v=(0,2^{2c},0)$,$v=(0,2^{2d},0)$. These data have been boxcar smoothed by a factor of n = 3 to a channel width of 0.82 $km\ s^{-1}$.}
\label{fig:sma2}
\end{figure}

Figure \ref{fig:sma1} shows the HCN (4-3) $v=(0,1^{1c},0)$, and $v=(0,1^{1d},0)$ spectra, while Figure \ref{fig:sma2} shows the HCN (4-3) $v = (0, 2^0, 0)$, $(0, 2^{2c}, 0)$ and $(0, 2^{2d}, 0)$ spectra. Table \ref{tab:table1} shows the measured line strengths, FWHM, and $v_{lsr}$ of the HCN transitions. The $v=0$ transition is the brightest with a temperature of $\sim39$ K. We observe the $v=1d$ line to be of approximately equal brightness to the $v=1c$ line. All three $v=2$ lines are of approximately equal brightness, with the exception of a single velocity component in the $v=2^{2c}$ line. All of the HCN emission lines show a v$_{lsr}=-5.3 \pm 0.7$ $km\ s^{-1}$ velocity shift.

The ratio of the measured brightness temperature of the three $v=2$ transitions to the two $v=1$ transitions can be used to estimate the excitation temperature of the gas in the optically thin limit. The ratio of the upper and lower state brightness temperatures are related to the excitation energy and excitation temperature for a linear molecule by:
\begin{equation}\label{eq:tmb}
\frac{T_{\text{MB},i}}{T_{\text{MB},l}}=\frac{\text{A}_{i}}{\text{A}_{l}}\left(\frac{\nu_{i}}{\nu_{l}}\right)^{2}\frac{g_{i}}{g_{l}}e^{-E_{il}/T_{\text{ex},il}}
\end{equation}
where $T_{\text{MB}_{i,l}}$ are the brightness temperatures (K) of the upper and lower states, A$_{i,l}$ are the Einstein coefficients, $\nu_{i,l}$ are the upper and lower frequencies,  g$_{i,l}$ are the upper and lower state degeneracies, E$_{il}$ is the excitation energy, and T$_{ex, il}$ is the excitation temperature. Averaging the excitation temperatures from all six $i=2$ to $l=1$ line ratios, we obtain an estimate of the excitation temperature to be 1050 $\pm$ 243~K. 

\cite{2007A&A...475..549B} have previously observed AFGL 2591 from 352.6 - 354.6 GHz at 0.6\arcsec\ spatial resolution in the \textit{extended} array configuration. Their observations cover the $v=0$ and ($v=1$) ro-vibrational transitions of HCN. Both groups find a velocity offset of $v_{lsr} = -5.3\pm0.7$ $km\ s^{-1}$, consistent with the systemic velocity of the molecular cloud surrounding AFGL 2591 \citep{1999ApJ...522..991V}. Spatially, we see the same extended  emission to the NW in the $v=0$ transition images as \cite{1999ApJ...522..991V,2006A&A...447.1011V} and \cite{2007A&A...475..549B}; however, our angular resolution is insufficient to resolve the two peaks. They measure a peak brightnesses of 2.3 Jy beam$^{-1} $and 3.1 Jy beam$^{-1}$, respectively. Our measurements of 24.4 and 6.6 Jy beam$^{-1}$ are $\sim11\times$ and $2.2\times$ larger than their measured values. Since \cite{2007A&A...475..549B} use a very different SMA configuration with much higher spatial resolution, comparison of the measured line fluxes is not straightforward. The large difference in measured flux for the $v=0$ HCN line suggests the emission is relatively extended and being resolved out in the work of \cite{2007A&A...475..549B}. Our larger beam recovers more extended flux. The difference between measurements of the $v=1$ line fluxes is smaller, suggesting marginal resolution of the line flux in \cite{2007A&A...475..549B}, but to a lesser degree than the $v=0$ line.

\section{Discussion}\label{section:discussion}

\subsection{Origin of Ro-Vibrational HCN Emission}\label{ro-vib}
Observations from the SMA of the $v=0$ and both $v=1$ lines are $\sim80\times$ and $90\times$ greater than the HHSMT measurements. For both lines, this is consistent with the $\sim70\times$ reduction in beam dilution provided by the SMA for a source unresolved by both telescopes.

Given the strong infrared radiation field surrounding AFGL 2591, one might expect any detected vibrationally excited HCN emission to be consistent with pumping through 14 $\mu$m absorption. In the SMA data, we observe the all three HCN (4-3) $v=2$ lines of approximately equal strength. Emission from the $v=(0, 2^{2c}, 0)$ and $v=(0, 2^{2d}, 0)$ lines is expected to be absent if purely radiative excitation dominates due to the $\Delta l = \pm1$ selection rule. The existence of lines from transitions forbidden to IR pumping at approximately the same brightness temperature as a permitted $v=(0,2^{0},0)$ line demonstrates that the excitation mechanism for these transitions must be primarily collisional. The small finger of emission at the $3\sigma$ level to the southwest in the $v = (0, 2^0, 0)$ line could be evidence of some extended emission resulting from infrared pumping. 

Figure \ref{fig:sma2} shows the $v=(0, 2^{2c}, 0)$ emission line which is almost twice as strong as the other $v=2$ lines, with excess emission primarily concentrated in a velocity component at -7.5 $km\ s^{-1}$. The source of this excess emission is unclear, and is inconsistent with the emission from the other two $v=2$ transitions. In a recent SMA study \citet{2007A&A...475..549B} note that the HCN (4-3) $v=1c$ line is preferentially emitted very close to the protostar ($\sim$600 AU) and, unlike other lines observed, its visibility diagram is not well described by models including far-UV and X-ray protostellar emission. The excess emission could be caused by maser activity which is seen in other transitions in the source \citep{2003ApJ...589..386T}, although the pumping mechanism is unclear.  Another possibility is irradiation by FUV/X-ray emission escaping the central source, enhancing the $v=(0,2^{2c},0)$ transition in the surrounding material \citep{2009A&A...503L..13B}. 

\subsection{Double Peaked Emission}\label{dd}
Both the $(0, 2^{0}, 0)$ and $(0, 2^{2d}, 0)$ spectral lines shown in Figure \ref{fig:sma2} appear to have double peaked emission with velocity components at -6.5 $km\ s^{-1}$ and -3.5 $km\ s^{-1}$. This could be the signature of Keplerian rotation. Construction of position-velocity diagrams are not possible because we do not have the spatial resolution necessary to resolve the disk. Instead, we measure the separation between the velocity components in the $v=(0, 2^{0}, 0)$ and $(0, 2^{2d}, 0)$ lines using gaussian fitting. This separation is $\sim2.7$ $km\ s^{-1}$.  Using Kepler's third law, assuming the central mass of AFGL 2591 to be $\sim10-16$ M$_{\odot}$, a disk inclination angle of $26 \pm 3^\circ$~\citep{2006A&A...447.1011V}, and the velocity centroid separation corresponds to the orbital velocity of the outermost disk edge, we estimate a disk diameter of $\sim1000-1800$ AU. This result is consistent with that of \cite{2006A&A...447.1011V} who estimate the source size morphologically using continuum emission measured with a synthesized beam an order of magnitude smaller than ours. The possibility also exists that this feature is self-absorption in the emission line, similar to the self absorption in the $v=0$ line reported by \cite{2007A&A...475..549B}, or result from other kinematic effects like infall. Observation the $v=2$ HCN (4-3) lines from this source with improved spatial resolution and sensitivity (i.e. with ALMA) will be able to clearly prove or disprove the Keplerian disk hypothesis by imaging the kinematic structure of the source. 

\begin{table*}[!htp]
      \caption{Results (errors) calculated from gaussian fit to the HCN spectral lines. }
   \label{tab:table1}
   \vskip 1pt
   \resizebox{0.8\textwidth}{!}{%
   \begin{tabular*}{\textwidth}{@{ }lccccc @{ }} 
      \cline{1-6}
     \multicolumn{6}{l}{}\\
      \cmidrule [2 pt]{1-6} 
Transition	&	Frequency	&	Peak $T_{B}$ SMA	&	FWHM	&	$V_{lsr}$	&	Filling Factor	\\
	&	GHz	&	K	&	$km\ s^{-1}$	&	$km\ s^{-1}$	&		\\
\cmidrule{1-6}
\textbf(HHSMT)	&		&		&		&		&		\\
(0,$1^{1c}$,0)	&	354.460433 (1.30E-05)	&	0.1080 (0.0150)	&	3.7000 (6.4000)	&	-5.3000 (3.0000)	&	--	\\
(0,0,0)	&	354.505473 (1.00E-06)	&	2.0300 (0.0150)	&	5.0000 (0.0700)	&	-5.6000 (0.0300)	&	--	\\
(0,$2^{2c}$,0)	&	356.135347 (5.00E-06)	&	$<0.0110$	&	--	&	--	&	--	\\
(0,$2^{2d}$,0)	&	356.162751 (5.00E-06)	&	$<0.0110$	&	--	&	--	&	--	\\
(0,$1^{1d}$,0)	&	356.255566 (1.30E-05)	&	0.017 (0.0140)	&	8.3000 (4.2000)	&	-11.4000 (2.5500)	&	--	\\
(0,$2^{0}$,0)	&	356.301176 (6.00E-06)	&	$<0.0110$	&	--	&	--	&	--	\\
\textbf(SMA)	&		&		&		&		&	\\	
(0,$1^{1c}$,0)	&	354.460433 (1.30E-05)	&	10.5240 (0.0305)	&	5.1201 (0.0114)	&	-5.1294 (0.0073)	&	0.0099 (2.8587E-05)	\\
(0,0,0)	&	354.505473 (1.00E-06)	&	38.9784 (0.1511)	&	6.6230 (0.0198)	&	-4.8467 (0.0126)	&	0.9164 (3.5524E-03)	\\
(0,$2^{2c}$,0)	&	356.135347 (5.00E-06)	&	1.9676 (0.0197)	&	5.4246 (0.0414)	&	-6.4622 (0.0263)	&	0.0009 (9.4025E-06)	\\
(0,$2^{2d}$,0)	&	356.162751 (5.00E-06)	&	0.8718 (0.0128)	&	6.6584 (0.0754)	&	-5.6232 (0.0477)	&	0.0004 (6.1093E-06)	\\
(0,$1^{1d}$,0)	&	356.255566 (1.30E-05)	&	12.2621 (0.0336)	&	5.1016 (0.0108)	&	-5.2010 (0.0079)	&	0.0115 (3.1487E-05)	\\
(0,$2^{0}$,0)	&	356.301176 (6.00E-06)	&	1.0692 (0.0149)	&	7.6095 (0.0847)	&	-4.4490 (0.0530)	&	0.0005 (7.1861E-06)	\\
      \cmidrule[2pt]{1-6}
   \end{tabular*}}%
\end{table*}

\subsection{Filling Factor}\label{ff}
We have calculated the filling factor of the SMA beam by dividing the brightness temperature of the spectral lines by the excitation temperature for the spectral feature (see Table \ref{tab:table1}). This assumes the brightness temperature of the emission region is $\sim T_{ex}$, true if those regions are optically thick. The range of filling factors is as expected, with the $v=0$ line the largest and the $v=2$ lines the smallest. We find the $v=0$ emission to approximately fill the beam. The filling factors of the $v=1$ lines are significantly larger than the $v=2$ lines, possibly due to shock heating or FUV irradiation of the outer envelope of HCN surrounding AFGL 2591 \citep{2009A&A...503L..13B}. The filling factors associated with the $v=2$ lines are approximately equal and very small, suggesting the emission is from compact unresolved sources within the beam, highlighting high density, warm structure in the disk and envelope.  

\section{Summary} \label {sum}
We have demonstrated the feasibility of a novel technique for the study of protostellar accretion disks using vibrationally excited HCN (4-3). We observed AFGL 2591 with the HHSMT and the SMA at 350 GHz. Selection rules for the $v=2$ HCN transitions allow us to determine the excitation mechanism independent of any other observations or assumptions. In AFGL 2591, the vibrationally excited HCN emission appears to be collisionally excited. The high densities and temperatures required for collisional excitation of these transitions imply that the emitting gas should trace structure in the protostellar envelope and/or disk surrounding the source. The line shapes of the $v=2$ lines suggest the possibility of of an unresolved, rotating Keplerian disk. Future followup with ALMA will be able to observe AFGL 2591 in the HCN (4-3) $v=2$ lines with sufficient sensitivity and spatial resolution to clearly detect or rule out the Keplerian disk we suggest. These lines seem to be promising probes of protostellar disk and envelope structure. The increased sensitivity and spatial resolution offered by ALMA will allow these tracers to be applied to a variety of protostellar sources, and should offer an improved way to isolate and study the very high density, warm regions around protostellar objects. 

\section{Acknowledgements}
The SMT is operated by the Arizona Radio Observatory (ARO), Steward Observatory, University of Arizona. The Submillimeter Array (SMA) is a joint project between the Smithsonian Astrophysical Observatory and the Academia Sinica Institute of Astronomy and Astrophysics and is funded by the Smithsonian Institution and the Academia Sinica.

\end{document}